\documentstyle[12pt]{article}
\newcommand{\be}{\begin{equation}}
\newcommand{\bea}{\begin{eqnarray}}
\newcommand{\eea}{\end{eqnarray}}
\newcommand{\ee}{\end{equation}}
\newcommand{\lb}{\label}
\begin{document}

\begin{titlepage}
\begin{flushright}
Freiburg THEP-94/20\\
gr-qc/9409014
\end{flushright}
\vskip 1cm
\begin{center}
{\large\bf  CONSISTENCY OF SEMICLASSICAL GRAVITY}
\vskip 1cm
{\bf Domenico Giulini and Claus Kiefer}
\vskip 0.4cm
 Fakult\"at f\"ur Physik, Universit\"at Freiburg,\\
  Hermann-Herder-Str. 3, D-79104 Freiburg, Germany.
\end{center}
\vskip 2cm
\begin{center}
{\bf Abstract}
\end{center}
\begin{quote}
We discuss some subtleties which arise in the semiclassical
approximation to quantum gravity. We show that integrability
conditions prevent the existence of Tomonaga-Schwinger time functions
on the space of three-metrics but admit them on superspace.
The concept of semiclassical time is carefully examined.
We point out that central charges in the matter sector spoil
the consistency of the semiclassical approximation unless the full
quantum theory of gravity and matter
 is anomaly-free. We finally discuss consequences
of these considerations for quantum field theory in flat spacetime,
but with arbitrary foliations.
 \end{quote}
\vskip 2cm
\begin{center}
{\em Submitted to Classical and Quantum Gravity}
\end{center}

\end{titlepage}

\section{Introduction}

Although a theory of quantum gravity is still elusive, formal
schemes have been developed, which are thought to exhibit some of the
important aspects of the full theory. One of these schemes is canonical
quantisation of general relativity and its central constraint equation
$H\Psi=0$. Although the special form of the Hamiltonian might well differ
from that of
 the fundamental theory, one would expect the constraint nature of
this equation to remain true, since this is a general feature of quantum
theories which are reparametrisation invariant at the classical level.

Given such a ``model theory", it is of central importance to recover
from the timeless form of the full constraint equation the limit where
spacetime can be considered as a classical object. There is general
agreement on the fact that a kind of Born-Oppenheimer approximation
with respect to the Planck scale plays a crucial r\^{o}le in this derivation
(see \cite{Ki94} for a review). In particular, one recovers an
approximate equation which is interpreted as a
 Tomonaga-Schwinger equation with
respect to a
``many-fingered time variable" $\tau(x)$ which controls the
dynamical behaviour of non-gravitational fields \cite{Ki94,Ze92}

The main purpose of the present paper is a critical investigation
into the meaning of this many-fingered time with regard to several
aspects. First, its very existence crucially depends on some
integrability conditions which, as it turns out, are not fulfilled
on the space of three-metrics,
Riem$\Sigma$ (where this time variable is usually written down),
but only on the space of three-geometries (``superspace").
Second, the possible presence of anomalies in the non-gravitational
sector prevents its existence even on superspace {\em unless}
the full quantum theory of gravity
and matter is anomaly-free. The consistency
of semiclassical gravity is thus already tight to the consistency of full
quantum gravity. Third, we shall show that semiclassical gravity can
only be interpreted in a sensible way if one considers, for any given
foliation of spacetime, a global Schr\"odinger equation with a global
time parameter and that it is very likely that one obtains unitarily
non-equivalent quantum theories for different foliations.

Our paper is organised as follows. Section~2 starts with a brief review
of the semiclassical expansion. We then discuss in detail the integrability
conditions for the existence of a time function. We show that these
conditions close on the diffeomorphism constraints and thus prevent
the existence of this time function on Riem$\Sigma$, but at the same
time guarantee its existence on superspace. The equations are explicitly
spelled out for the case of a scalar field.

In section~3 we point out that anomalies in the commutator between
the matter Hamiltonian and the matter momentum
 might spoil the integrability conditions even
on superspace unless full quantum gravity is anomaly-free. We illustrate
this point by a two-dimensional example, the Virasoro algebra, which
is well known from string theory.

In the last section we present a brief summary and discuss the general
meaning of our results for a diffeomorphism invariant quantum theory.

\section{The r\^{o}le of the diffeomorphism constraints in the semiclassical
         expansion}

The basic idea in the semiclassical expansion is to start from the
full Hamiltonian constraint equation in quantum general relativity and
to perform an expansion in powers of the gravitational constant.
In the geometrodynamical approach, in particular, the starting point is the
Wheeler-DeWitt equation (we set $c=1$)
\be {\cal H}\Psi[h_{ab},\phi] \equiv
    \left(-16\pi G\hbar^2 G_{abcd}
    \frac{\delta^2}{\delta h_{ab}\delta h_{cd}}
    -\frac{\sqrt{h}}{16\pi G}(R-2\Lambda)
    +{\cal H}_m\right)\Psi=0, \lb{1} \ee
where
$h_{ab}$ is the three-metric, $R$ the three-dimensional Ricci scalar,
$\Lambda$ the cosmological constant, $G_{abcd}$ the contravariant
DeWitt metric,
and ${\cal H}_m$ is the Hamiltonian density for non-gravitational
fields.
We emphasise that (\ref{1}) is actually an {\em infinity} of equations,
one at each space point. More precisely, they should be understood
to be distributional equations which have to be integrated against
``lapse functions", $N(x)$, from a suitably chosen space of test functions.
This distinction is important when one considers open three-manifolds,
since then surface terms appear and one can no longer write the equation
in the local form (\ref{1}). In this paper, however, we restrict
attention to the closed case. The integrated form of (\ref{1}) reads
\be \int d^3x\ N{\cal H}\Psi \equiv {\cal H}^N\Psi
    \equiv ({\cal H}^N_G +{\cal H}^N_m)\Psi=0, \lb{2} \ee
where we have decomposed the full Hamiltonian into its gravitational
and matter parts, respectively. The ansatz
\be \Psi=\exp\left(i(MS_0+S_1+M^{-1}S_2+\ldots)/\hbar\right), \lb{3} \ee
where $M\equiv (32\pi G)^{-1}$, then leads, when inserted into (\ref{1}),
to a set of equations at consecutive orders in $M$.

The highest order, $M^2$, yields that $S_0$ must only depend on the
three-metric. The next order, $M$, gives the Hamilton-Jacobi equation
for the gravitational field,
\be {\cal H}_x= \frac{1}{2}G_{abcd}
    \frac{\delta S_0}{\delta h_{ab}}\frac{\delta S_0}{\delta h_{cd}}
     -2\sqrt{h}(R-2\Lambda) =0. \lb{4} \ee
Again, these are infinitely many equations and should be interpreted
in an integrated version as providing {\em one} equation for
{\em each} lapse function $N(x)$ out of the specified class of
test functions. The Hamilton-Jacobi equations (\ref{4}) are, together
with the momentum constraints in this order (see below), equivalent
to all Einstein field equations, and a family of classical spacetimes
can be constructed from a solution $S_0$ to the infinitely many equations
(\ref{4}).

At the next order, $M^0$, it turns out to be convenient to introduce a
functional
\be \psi\equiv D[h_{ab}]\exp(iS_1/\hbar) \lb{5} \ee
and demand that $D$ obeys
\be G_{abcd}\frac{\delta S_0}{\delta h_{ab}}\frac{\delta D}{\delta h_{cd}}
    -\frac{1}{2}G_{abcd}\frac{\delta^2 S_0}{\delta h_{ab}\delta h_{cd}}
    D= 0. \lb{6} \ee
This corresponds to the usual equation for the WKB prefactor.
The important observation at this order is that $\psi$ obeys the equation
\be i\hbar G_{abcd}\frac{\delta S_0}{\delta h_{ab}}
    \frac{\delta\psi}{\delta h_{cd}} ={\cal H}_m\psi. \lb{7} \ee
The left-hand side has the interpretation of being $i\hbar$ times the
derivative along the vector fields (one vector field for each space point)
\be \chi(x) = G_{abcd}(x)\frac{\delta S_0}{\delta h_{ab}(x)}
    \frac{\delta}{\delta h_{cd}(x)} =
    -2K_{ab}(x)\frac{\delta}{\delta h_{ab}(x)}, \lb{8} \ee
where
\[ K_{cd}= -\frac{1}{2}G_{abcd}\frac{\delta S_0}{\delta h_{ab}}, \]
which in the classical spacetime constructed from $S_0$ has the interpretation
of the extrinsic curvature. Note, however, that $K_{ab}$ is here
considered as a functional of the three-metric.

If we were allowed to write
\[ \chi(x)=\frac{\delta}{\delta\tau(x)}, \]
Eq. (\ref{7}) would just be a Tomonaga-Schwinger equation with respect to
the many-fingered time functionals $\tau(x)$. Note that $\tau$
is really a time functional on Riem$\Sigma$ {\em for each x}.
It would thus be more precise to write, as is often done,
$\tau[h_{ab},x)$, but we shall not use this notation explicitly.

However, if we used such time functionals in (\ref{7}), this would lead
us immediately to a contradiction: The commutator $[\delta/\delta
\tau(x),\delta/\delta\tau(y)]$ necessarily vanishes
but the commutator $[{\cal H}_m(x),\ {\cal H}_m(y)]$
arising from the right-hand side does not vanish -- it closes on the
momentum density of the matter field. We emphasise that this
problem is different from the ``global time problem" \cite{Ku92}
which prevents the
{\em global} existence of such a time variable in configuration
space.

To simplify the discussion we write (\ref{7}) in the form
\be i\hbar\chi^N\psi ={\cal H}_m^N\psi, \lb{9} \ee
where we have introduced the vector fields $\chi^N$ (giving {\em one}
vector field for {\em each} choice of lapse) by integrating (\ref{8})
with respect to $N$, and ${\cal H}_m^N$ was defined above, see
Eq. (\ref{2}). It is clear that
$[\chi^N,\chi^M]=0$ is a necessary condition for the local existence
of time functionals $\tau^N$ such that $\delta/\delta\tau^N =
\chi^N$.

It is straightforward to calculate the commutator
\bea [\chi^N,\chi^M] &=& \left[\int_x N(x)G_{abcd}(x)
     \frac{\delta S_0}{\delta h_{ab}(x)}\frac{\delta}
     {\delta h_{cd}(x)}\right., \nonumber\\
     & & \ \left.\int_y M(y)G_{nmrs}(y)\frac{\delta S_0}{\delta h_{nm}(y)}
     \frac{\delta}{\delta h_{rs}(y)}\right] \nonumber\\
     &=& \int_x\int_y (N(x)M(y)-M(x)N(y))G_{abcd}(x)\frac{\delta S_0}
     {\delta h_{cd}(x)} \times \nonumber\\
     & & \ \frac{\delta^2 S_0}{\delta h_{rs}(y)\delta h_{ab}(x)}
     \int_z\delta(y,z)G_{rsnm}(z)\frac{\delta}{\delta h_{nm}(z)},
     \lb{10} \eea
which vanishes only if
\[     \int_x (N(x)M(y)-M(x)N(y))G_{abcd}(x)\frac{\delta S_0}
     {\delta h_{cd}(x)}
      \frac{\delta^2 S_0}{\delta h_{rs}(y)\delta h_{ab}(x)} \]
vanishes for {\em all} $N$ and $M$. This is equivalent to the condition
 \be     \frac{\delta^2 S_0}{\delta h_{ab}(x)\delta h_{cd}(y)}
         \propto \delta(x,y). \lb{11} \ee
We shall, however, show that this cannot occur. Differentiating
the Hamilton-Jacobi equations (\ref{4}) with respect to $h_{ab}(y)$,
one obtains
\be \frac{\delta{\cal H}_x}{\delta h_{ab}(y)}
 +\int_z \frac{\delta{\cal H}_x}{\delta \pi^{cd}(z)}
      \frac{\delta^2 S_0}{\delta h_{cd}(z)\delta h_{ab}(y)}=0,
      \lb{12} \ee
where we have written $\pi^{cd}(z)=\delta S_0/\delta h_{cd}(z)$
(this is $32\pi G$ times the classical geometrodynamical momentum).
The first term in (\ref{12}) captures the explicit dependence
of ${\cal H}_x$ on the three-metric, while
the second term takes into account the implicit dependence
through the replacement $\pi^{ab} \to\delta S_0/\delta h_{ab}$.

{}From (\ref{4}) and(\ref{12}) one has
\be \frac{\delta{\cal H}_x}{\delta h_{ab}(y)}=
    -G_{cdnm}(x)\frac{\delta S_0}{\delta h_{nm}(x)}
      \frac{\delta^2 S_0}{\delta h_{cd}(x)\delta h_{ab}(y)}. \lb{13} \ee
The derivative on the left-hand side is given by the expression
\be \frac{\delta{\cal H}_x}{\delta h_{ab}(y)}=
    2G^{abcd}(y)\delta_{\vert cd}(x,y) +F^{ab}(x)\delta(x,y), \lb{14} \ee
where the explicit form of the second term will be
irrelevant due to its ultralocal form (no derivatives of the
delta function).
 We note that the second derivative
of the delta function occurs from differentiating the Ricci scalar.
Inserting (\ref{13}) and (\ref{14}) into (\ref{10}) then yields
\be [\chi^N,\chi^M] =-2\int_x\int_y (N(x)M(y)-M(x)N(y))
    \delta_{\vert ab}(x,y)\frac{\delta}{\delta h_{ab}(y)}, \lb{15} \ee
which after some partial integrations can be written as
\bea [\chi^N,\chi^M] &=& -2\int_x (N\partial_a M-M\partial_a N)
     \left(\frac{\delta}{\delta h_{ab}}\right)_{\vert b}
     \nonumber\\
     &=& \int{\cal L}_{{\bf K}}h_{ab}\frac{\delta}{\delta h_{ab}},
      \lb{16} \eea
where
\[ K^a=h^{ab}(N\partial_bM -M\partial_bN). \]
Thus, $[\chi^N,\chi^M]\neq 0$, and time functions in the above sense can
{\em never} be introduced. Formally, this happens because the Ricci scalar
$R$ is not ultralocal in $h_{ab}$, which leads to the occurrence of the
second derivative of the $\delta$ distribution in (\ref{14}).
It is, however, not surprising that the commutator closes on the
generator of a diffeomorphism -- the vector fields $\chi^N$ which appear
in the semiclassical approximation are the generators of a hypersurface
deformation normal to itself, whose commutator is known to
generate deformations tangential (``stretchings") to the
hypersurface \cite{HKT}.

In the De Sitter example discussed in \cite{Ki94} it was found that
$S_0\propto \int d^3x\sqrt{h}$ which obeys relation (\ref{11}).
This is not in contradiction with the discussion above, where we
assumed $S_0$ to be a solution to the full Hamilton-Jacobi
equation. In the de~Sitter example the expression
for $S_0$ only solves the Hamilton-Jacobi equation on the
submanifold of $\mbox{Riem}\Sigma$ where the Ricci scalar vanishes.
For De~Sitter space this can always be achieved by a specific
foliation.

A proper understanding of (\ref{16}) and its compatibility with the
semiclassical equations (\ref{7}) is obtained if one expands,
in addition to the Wheeler-DeWitt equation (\ref{1}), the
diffeomorphism constraints in powers of $G$. These are given by
\be 2h_{bc}D_a\frac{\delta\Psi}{\delta h_{ab}}
    =\phi_{,c}\frac{\delta\Psi}{\delta\phi}, \lb{17} \ee
where we have, for simplicity, chosen a single scalar field for
the non-gravitational sector. We emphasise that the gravitational
constant is {\em absent} from this equation, which renders its
expansion fairly trivial.

The highest order, $M$, yields (since $S_0$ does not depend on $\phi$)
\be 2h_{bc}D_a\frac{\delta S_0}{\delta h_{ab}}=0. \lb{18} \ee
This expresses nothing but the diffeomorphism invariance of the solutions,
$S_0$, to the Hamilton-Jacobi equations (\ref{4}).

The next order, $M^0$, leads to a condition on the functional $\psi$,
Eq. (\ref{5}),
\be 2h_{bc}D_a\left(\frac{\delta\psi}{\delta h_{ab}}
    -\frac{\psi}{D}\frac{\delta D}{\delta h_{ab}}\right)
    =\phi_{,c}\frac{\delta\psi}{\delta\phi}. \lb{19} \ee
Since the prefactor $D$ depends only on the three-metric, see (\ref{6}),
it is appropriate to demand that it be diffeomorphism invariant by
itself, i.e.,
\be h_{bc}D_a\frac{\delta D}{\delta h_{ab}}=0. \lb{20} \ee
{}From (\ref{19}) one thus finds
\be 2h_{bc}D_a\frac{\delta\psi}{\delta h_{ab}}= \phi_{,c}
    \frac{\delta\psi}{\delta\phi} \lb{21} \ee
which is of the same form as the general equation (\ref{17}).
Thus, it expresses the invariance of the wave functional $\psi
[h_{ab},\phi]$ with respect to simultaneous diffeomorphism of the metric
and the matter field. The consistency condition for (\ref{9}) reads
\be [\chi^N,\chi^M]\psi =[{\cal H}_m^M,{\cal H}_m^N]\psi. \lb{22} \ee
This, however, is nothing but the momentum constraint in this order of
approximation, Eq. (\ref{21}), since $[\chi^N,\chi^M]$
generates a diffeomorphism of the metric, Eq. (\ref{16}), and
$[{\cal H}_m^M,{\cal H}_m^N]$ closes on the momentum density of matter
which generates a diffeomorphism of the matter field.
Thus, as in the full theory \cite{MT}, the momentum constraints
provide the {\em integrability conditions} for the Tomonaga-Schwinger
equations (\ref{9}).

In the explicit case of a scalar field one has, for example,
\be [{\cal H}_m^M,{\cal H}_m^N]= -\int_x (N\partial_aM
    -M\partial_aN)h^{ab}\phi_{,b}\frac{\delta}{\delta\phi},
    \lb{23} \ee
which, together with (\ref{16}), yields (\ref{21}).

Although a family of time functions $\tau(x)$ on Riem$\Sigma$
does not exist, one can integrate (\ref{9}) along the vector field
$\chi^N$ {\em for one particular choice of lapse} $N$.
This defines a {\em global} time parameter $t$ with respect to which
{\em one} global Schr\"odinger equation can be written down.
It is in this sense that quantum field theory with respect to
a chosen foliation emerges from full quantum gravity.
Such a picture has been implicitly used by many authors although
 (non-existing) time functions $\tau(x)$ have been used \cite{Ki94}.
One may then proceed to the next order of approximation and derive
correction terms to the functional Schr\"odinger equation
\cite{Ki94,KS}. Again, these terms have to be interpreted with respect
to a particular foliation.

We emphasise that the same criticism applies to the standard
Tomonaga-Schwinger equations in flat spacetime since the matter
Hamiltonians always close on the momentum density. One should thus be
careful and consider these equations only in the context of a parametrised
formalism where all possible embeddings of a spacelike hypersurface
into spacetime are allowed. This leads to the same interpretation
as above. One would, however, not expect that the resulting
quantum theories are foliation independent.

If there are no such time functions on Riem$\Sigma$, can there
be such functions on superspace $S(\Sigma)\equiv \mbox{Riem}\Sigma
/\mbox{Diff}\Sigma$? To answer this question we must project the vector
fields $\chi^N$ down on superspace. This is possible since $\chi^N$
is invariant under diffeomorphisms -- a property which it inherits from
the diffeomorphism invariance of $S_0$ and $\int N(x)G_{abcd}(x,x')$.
Moreover, $\chi^N$ is (with respect to DeWitt's metric)
orthogonal to the diffeomorphism orbits and thus can be projected
onto non-vanishing vector fields $\bar{\chi}^N$ in specifiable
regions of superspace \cite{Gi94}: $\chi^N \stackrel{\pi_{*}}
{\to}\bar{\chi}^N$. One thus has
\[ \pi_*[\chi^N,\chi^M]=[\pi_*\chi^N,\pi_*\chi^M]
   =[\bar{\chi}^N,\bar{\chi}^M]=0. \]
Thus, there exist functions $\bar{\tau}^N$ on superspace such that
\[ \bar{\chi}^N \equiv \frac{\delta}{\delta\bar{\tau}^N}. \]
Since the Wheeler-DeWitt operator is, however, only defined on
Riem$\Sigma$ and not on $S(\Sigma)$ (an object such as the second
derivative with respect to three-geometry does not exist
globally), the physical
interpretation of these time functions is exhibited only implicitly
by first performing the calculations on Riem$\Sigma$ and then projecting
on superspace.

We finally note that the same discussion applies to the semiclassical
approximation of connection dynamics \cite{Ki94}. One must,
however, take care for the fact that the commutator between
two Hamiltonians does not close on the diffeomorphism constraint,
but contains an additional contribution from the gauge sector.

\section{Anomalies}

The ``smeared out version" of the momentum constraints (\ref{21})
reads
\be {\cal H}_G^{{\bf N}}\psi ={\cal H}_m^{{\bf N}}\psi, \lb{24} \ee
where ${\bf N}(x)$ denotes the shift vector field. This implies,
together with (\ref{9}), the consistency condition
\be  i\hbar [\chi^N,{\cal H}_G^{{\bf N}}]\psi
    =[{\cal H}_m^{{\bf N}},{\cal H}_m^N]\psi. \lb{25} \ee
Classically, the commutator on the right-hand side closes again
on the matter Hamiltonian density ${\cal H}_m^M$, where
$M={\cal L}_{\bf N}N$. In the quantum theory, however, one knows that
-- under certain assumptions (see below) -- the commutator on the
right-hand side {\em necessarily} leads to central extensions
(Schwinger terms, anomalies) \cite{BD}. It thus seems that the semiclassical
expansion is inconsistent unless this anomaly is cancelled by a similar
anomaly occurring on the left-hand side. It is, however, straightforward
to show that this cannot happen. Basically, the reason is that $\chi^N$
is only the {\em classical} generator of hypersurface deformations
and thus contains, as well as ${\cal H}_G^{\bf N}$, only {\em single}
derivatives with respect to the metric. Therefore, the left-hand side
of (\ref{25}) yields the classical result and closes on a hypersurface
deformation:
\bea & & [\chi^N,{\cal H}_G^{\bf N}]
     = \left[\int_x NG_{abcd}\frac{\delta S_0}{\delta h_{ab}}
     \frac{\delta}{\delta h_{cd}},-2\int_y N^kh_{lk}D_m
     \frac{\delta}{\delta h_{ml}}\right] \nonumber\\
     &=& \int_x M(x)G_{abcd}\frac{\delta S_0}{\delta h_{ab}}
     \frac{\delta}{\delta h_{cd}} =\chi^M, \lb{26} \eea
where $M={\cal L}_{\bf N}N$. In the absence of an anomaly the condition
(\ref{25}) would thus again lead to (\ref{9}).

At this point we recall that the necessary occurrence of anomalies
was only shown under the following conditions \cite{BD}:
(1) Locality, (2) the existence of a ground state for ${\cal H}_m$
(with $N=constant$), Lorentz invariance, and the existence of a
positive definite Hilbert space product. But of these assumptions,
Lorentz invariance is definitely an inappropriate assumption
from the viewpoint of full quantum gravity,
while the existence of a ground state is at least dubious.
If the regularisation of the full theory can be performed in a
diffeomorphism invariant manner (otherwise one would interpret the
full theory as being inconsistent \cite{CJ}), anomalies
should be absent in the full theory and, consequently, also in the
semiclassical consistency equation (\ref{25}). If one nevertheless
insists on a Lorentz covariant regularisation of the right-hand side
of (\ref{25}), anomalies appear, reflecting the inappropriateness
of the regularisation procedure as viewed from the fundamental
theory. Under the assumptions made in \cite{BD} the anomaly must be
proportional to $\delta^{'''}(x-y)$, where the constant of proportionality
is given, in three space dimensions, by an inverse length squared,
exhibiting an ultraviolet singularity as one approaches
$L\to 0$. This anomaly would then have to be cancelled by a
similar anomaly coming from the {\em full} gravitational commutator
$[{\cal H}_G^N,{\cal H}_G^{\bf N}]$. The latter must, by dimensional
arguments, be proportional to $G^{-1}\delta^{'''}(x-y)$ (recall that
the gravitational constant, $G$, already appears, in contrast to the
classical case, in pure quantum gravity). In other words, $G$ would then
play the r\^{o}le of a ``regulator" for ultraviolet divergencies.
The total, regularised, commutator
 \[ [{\cal H}_G^N,{\cal H}_G^{\bf N}]
_{reg}\Psi =
[{\cal H}_m^{\bf N},{\cal H}_m^N]_{reg}\Psi\]
 would then
exhibit no anomalies, and (\ref{25}) would be consistent.

While little is known in four spacetime dimensions, the situation
in two dimensions has been studied extensively, mostly in the context of
string theories \cite{GSW}. There, the central extension in
$[{\cal H}_m^{\bf N},{\cal H}_m^N]$ is given by the central charge
of the Virasoro algebra. Its appearance is usually found from a normal
ordering prescription for the Hamiltonian. This, of course, does not mean
that two-dimensional quantum gravity is inconsistent. The full theory
can be regularised in a diffeomorphism invariant way but {\em not}
in a way which respects both Weyl invariance and conformal invariance
(this happens only in the critical string dimension). If $\Gamma
[g_{\mu\nu}]$ denotes the effective action which arises from integrating out
matter in the full action $S[g_{\mu\nu},\phi]$, one has
(see, e.g., \cite{SWD})
\be 0= \delta\Gamma_{Diff}=\delta\Gamma_{conf} -\delta\Gamma_{Weyl},
      \lb{27} \ee
where the quantities on the right-hand side denote the change of
the effective action with respect to conformal transformations
and Weyl transformations, respectively. Both changes are, of course,
related to the central charge, and the appearance of a Schwinger
term in the Virasoro algebra is, in a sense, a consequence of the
diffeomorphism invariance of the full theory. That the trace anomaly
appears as a condition for the full Schwinger term to vanish has been
explicitly studied in a two-dimensional model by Teitelboim \cite{Te83}.

Kucha\v{r} \cite{Ku89} has cast the theory of a massless free scalar
field in two spacetime dimensions into a parametrised form
(this mimics the diffeomorphism invariance of quantum gravity).
He could show that this enabled him to introduce an embedding-
dependent anomaly potential which cancels the anomaly in the
Dirac algebra. He argued that, provided certain topological
conditions on the space of embeddings are met, the anomaly can be
absorbed into a redefinition of momenta. These topological
conditions are met in two dimensions, but not necessarily in
four.\footnote{The anomaly appears formally as a closed
two-form on the manifold of surface embeddings. A sufficient
topological condition for it to be exact is that the second
cohomology vanishes (and not, as stated in \cite{Ku89},
the first homotopy).}

The two-dimensional case clearly demonstrates the dependence of anomalies
on the regularisation prescription. One can, for example, consider
a whole family of possible ground states with respect to which
a normal ordering prescription can be performed \cite{NS}.
Consequently, the obtained central charge depends on the special
ground state chosen from this family. One can in this case
make a choice such that the anomaly cancels against a similar
anomaly from the ghost sector. This drastically illustrates how
sensibly the concept of anomaly depends on the method of
``quantisation".

In summary, we have shown that the semiclassical expansion can
be carried out consistently if the full theory does not
possess central charges.

\section{Conclusion}

One of the important lessons to be drawn from the above considerations
is in our opinion the fact that an approximate theory contains
indications which point towards the more fundamental theory. In our
case this happens through the occurrence of anomalies if a subsector
of the theory is regularised with respect to approximate concepts:
If the functional Schr\"odinger equation is approximately valid,
and if the background gravitational field is very weak, it is very
suggesting to take Lorentz invariance as a guiding principle for
the approximate theory. Then anomalies necessarily
occur which would lead to an inconsistency if one attempted to
write down a Tomonaga-Schwinger equation. Full quantum gravity, however,
is supposed to obey the more general diffeomorphism invariance
which should allow a consistent regularisation of the theory
and should thus be anomaly-free. Only if one imposed a restricted
regularisation principle on subsectors of the theory (such
as the matter fields alone) could one find anomalies which must then
be cancelled by similar anomalies from the remaining part of the full
theory. This is clearly illustrated in the two-dimensional theory.

Another important point concerns the interpretation of the
Tomonaga - Schwing\-er equations in flat spacetime. As we have shown, even
there one cannot introduce time functions $\tau(x)$ on the space of
embeddings. One can only write down global Schr\"odinger equations
for {\em given} foliations and a corresponding global time
parameter. However, one would not expect that quantum field theories
corresponding to different foliations would
in general be unitarily equivalent.
The reason for this expectation is the fact that particle number
is not a diffeomorphism invariant quantity. If one produced, for example,
a local ``bump" in the hypersurfaces through a rapidly varying lapse
in a local region, this would correspond to a strong gravitational
field which would lead to strong particle creation and thus to
a Fock space which is orthogonal to the standard Fock space referring
to a flat foliation. Unitary equivalence would only be expected
for restricted classes of foliations such as the ones corresponding
to inertial observers.

Quite generally, symmetry groups which allow for central extensions
and therefore anomalous commutators cease to do so if embedded
in a larger group of symmetries. An effective reduction of the
symmetries might therefore lead to spurious anomalies.

\vskip5mm

\begin{center}
{\bf Acknowledgements}
\end{center}
We are most grateful to Roman Jackiw for correspondence and
 for motivating us to
study these problems. We also thank
 Petr H\'{a}j\'{\i}\v{c}ek and Andreas Wipf
 for useful discussions and critical comments.


\end{document}